\documentclass[12pt,preprint]{aastex}

\begin{document}
\title{Will Jets Identify the Progenitors of Type Ia Supernovae?}

\author{Mario Livio, Adam Riess {\&} William Sparks}

\affil{Space Telescope Science Institute, 3700 San Martin Drive, Baltimore, MD 21218}

\begin{abstract}
We use the fact that a Type~Ia supernova has been serendipitously discovered near the jet of the active galaxy 3C~78 to examine the question of whether jets can enhance accretion onto white dwarfs. One interesting outcome of such a jet-induced accretion process is an enhanced rate of novae in the vicinity of jets. We present results of observations of the jet in M87 which appear to have indeed discovered 11 novae in close proximity to the jet. We show that a confirmation of the relation between jets and novae and Type~Ia supernovae can finally identify the elusive progenitors of Type~Ia supernovae.
\end{abstract}

\section{Introduction}
Type Ia Supernovae (SNe Ia) are believed to represent complete thermonuclear disruptions of mass accreting white dwarfs at the Chandrasekhar limit \citep[e.g.][]{hoyle60,livio01,renzini96,nomoto00}. They are the most luminous type of stellar death, rivaling the optical luminosity of their host galaxies for a few weeks, and have been detected at up to a redshift $z\sim1.7$, when the universe was approximately 
one-quarter of its present age. SNe~Ia are also the most homogeneous of all supernova types, elevating them in recent years to the tool of choice for measuring the history of cosmic expansion. In 1998, two teams observed independent samples of SNe~Ia at redshifts $z\gtrsim0.3$ and concluded that the expansion of the universe appears to be accelerating, perhaps propelled by a kind of vacuum (``dark'') energy once posited then rejected by Einstein \citep{riess98,perlmutter98}. Studies are underway to use expanded samples of SNe~Ia to ferret out the nature of this dark energy. A serendipitous set of observations by the Hubble Space Telescope (HST) in 1997--1998 of the most distant 
SN~Ia (SN~1997ff) provided evidence for a period of cosmic deceleration, preceding today's acceleration, and caused by the gravitational influence of dark matter \citep{riess01}. Despite these recent successes in precision cosmological cartography from SNe~Ia, an embarrassing ignorance haunts the use of SNe Ia: what are their progenitors? Specifically, how does a sub-Chandrasekhar mass white dwarf grow to the critical mass? Is matter slowly accreted onto the white dwarf from a normal companion star or does the extra mass come all at once from a merger with another white dwarf? While theoretical models abound (see e.g.\ Livio 2001 and Hillebrandt \& Niemeyer 2000 for reviews), to date there has been no observational evidence for the accretion process, and the nature of the progenitor systems of SNe Ia has remained elusive. In the present Letter we show that another set of serendipitous observations by HST may have provided an important clue. 

\section{Serendipity and SNe Ia}
On September 6, 2000, the Space Telescope Imaging Spectrograph (STIS) on board HST took a 10~second image of a tiny region ($5''$ across) around the nucleus of the radio-loud galaxy 3C~78. The goal of this observation was to better pinpoint the location of the nucleus before the spectrograph would obtain its spectrum \citep{martel02}. The short acquisition image (see Figure~1) revealed a stellar source superimposed very near the jet and of similar brightness as the galactic nucleus. The subsequent spectroscopy taken of the nucleus and jet, serendipitously included the stellar source and provided 
the unambiguous spectral signature of a SN~Ia approximately 3~weeks after maximum brightness. This SN~Ia, designated SN~2000fs, was at the same redshift as the host galaxy ($z=0.0286$) and its apparent brightness was also that expected for a 3-week old SN~Ia.

Only about a dozen active galactic nuclei (AGN) have been observed with optical jets (as opposed to radio jets which are ubiquitous in radio galaxies). SNe~Ia are also rare, with only one expected in a galaxy (of Milky Way size) every 300 years \citep[e.g.][]{cappellaro93}. The suggestive significance of the superposition of these two rarities cannot be properly evaluated until a designed study of circumnuclear transients in AGN is undertaken. However, notwithstanding the recognized dangers of a posteriori probabilities, it is interesting to examine the question of whether a causal connection between the jet and the SN may be anticipated. 

Theoretical models suggest that SNe~Ia arise from central carbon ignition in accreting carbon-oxygen white dwarfs, as the latter reach the Chandrasekhar limit, having accreted at a rate \citep[e.g.][]{nomoto91}
10$^{-8}\lesssim\dot{M}_\mathrm{acc}\lesssim10^{-6}$ M$_{\odot}$~yr$^{-1}$. 
Typically, these models assume that the mass is provided by a companion star, because the accretion rate from the interstellar medium (ISM) is too low. Jets can change this paradigm in two important ways: (i)~The compressive effect of the shock waves produced by the jet may form denser clouds in the ISM \citep{woodward76,elmegreen78}, and 
(ii)~mass entrainment in the mixing layer of the jet can transport parcels of the ISM to regions removed from the normal extent of the stellar populations of the galaxy \citep{deyoung97}. Concerning the first effect ((i) above), it has even been suggested that jets can trigger star formation \citep[e.g.][]{deyoung81,vanbreugel85,rees89,begelman89,chambers90,daly90,bicknell00}. Evidence for the second effect ((ii) above) has been seen through the detection (e.g.\ in M87) of optical emission line regions surrounding jets and extended radio lobes \citep{sparks93}. 

The speed with which the bow shock (of cross-section~A) produced by the jet propagates into a cloud (of density $n$) is given approximately by 
\begin{equation}
v_\mathrm{sh}\simeq1000\eta
\left( \frac{v_\mathrm{jet}}{c} \right)^{-\frac{1}{2}}
\left( \frac{F}{10^{46}~\mathrm{erg~s}^{-1}} \right)^{\frac{1}{2}}
\left( \frac{A}{10^{43}~\mathrm{cm}^2} \right)^{-\frac{1}{2}}
\left( \frac{n}{\mathrm{cm}^{-3}} \right)^{-\frac{1}{2}}~\mathrm{km~s}^{-1}~~,
\end{equation}
where $F$ is the jet's energy flux, $v_\mathrm{jet}$ is its speed and $\eta$ is a dimensionless number of order unity which distinguishes between relativistic and nonrelativistic jets. Typical compression times are of order \citep[e.g.][]{elmegreen78}
\begin{equation}
\tau_\mathrm{comp}\simeq5\times10^6
\left( \frac{\xi}{100}\right)^{-\frac{1}{2}}
\left( \frac{n}{\mathrm{cm}^{-3}}\right)^{-\frac{1}{2}}~\mathrm{yr}~~,
\end{equation}
where $\xi$ is the compression factor (of order ${\cal M}^2$ in radiative shocks, where ${\cal M}$ is the Mach number of the shock). The size of the compressed region ($\sim\xi^{-1} v_\mathrm{sh} \tau_\mathrm{comp}$) is therefore of the order of a few hundred parsecs after a few million years. A compressed clump will cool on a timescale of
\begin{equation}
\tau_\mathrm{cool}\sim3\times10^5
\left( \frac{v_\mathrm{sh}}{1000~\mathrm{km~s}^{-1}} \right)
\left( \frac{n_\mathrm{cl}}{100~\mathrm{cm}^{-3}} \right)^{-1}~~\mathrm{yr}~~,
\end{equation}
where $n_\mathrm{cl}$ is the clump's particle density.

A white dwarf in such a dense ISM clump is expected to accrete at a rate of \citep{bondi52}
\begin{equation}
\dot{M}_\mathrm{acc}\simeq1.7\times10^{-9}
\left( \frac{M_\mathrm{WD}}{\mathrm M_{\odot}} \right)^2
\left( \frac{n_\mathrm{cl}}{100~\mathrm{cm}^{-3}} \right)
\left( \frac{T}{80~\mathrm{K}} \right)^{-\frac{3}{2}}~\mathrm{M_{\odot}~yr^{-1}}~~,
\end{equation}
where $M_\mathrm{WD}$ is the white dwarf's mass and $T$ is the clump's temperature. Type~Ia supernovae therefore could occur in clumps in which compression increased the density by at least an order of magnitude. 
Equation~(3), however, has another interesting consequence. Classical novae are obtained when white dwarfs accrete at rates satisfying \citep{prialnik95} 
$\dot{M}_\mathrm{acc}\lesssim10^{-8}$ M$_{\odot}$~yr$^{-1}$. If a jet-induced accretion process truly occurs as suggested above, this would imply that \textit{an enhanced rate of classical novae should also be observed around AGN jets}. Another unexpected observational result suggests that this may indeed be the case.

\section{Serendipity and Classical Novae}
An HST program intended to measure the proper motion of the optical jet in the active galaxy M87 revealed despite a small $7''$ field of view a surprising number (at least eleven) of transient stellar-like sources in the vicinity of the jet \citep{sparks00},  with two being very near the jet (Fig.~2). The $U$~magnitudes of these sources are in the range 22.6 to 25.8 (except for the brightest, to be discussed shortly). For a distance modulus to Virgo of 31.0, the implied absolute $U$~magnitudes are of order $-5$ to $-10$, those expected for classical novae.  Since the six epochs of observations were approximately one year apart, the implied duration of each transient event is of the order of months. Thus, the most likely interpretation, based on both their apparent luminosities and the timescale on which they vary, is that these sources are classical novae. A preliminary statistical estimate based on the apparent nova rate near the jet suggests an enhanced rate that could be a factor of 2--10 above the average rate 
\citep[200--1000 novae per year compared to the estimated 90--160;][]{dellavalle94,shafter00}. Since both observations and theory have previously demonstrated that classical novae result from gradual accretion onto white dwarfs \citep[e.g.][]{starrfield92,truran82,warner95}, the potential association of an unexpectedly large number of classical novae with the jet in M87 suggests that the jet may facilitate such accretion. We should note that while the currently available data are not sufficient to draw any definitive conclusions, they do suggest a preference for the transients to lie along the jet. Interestingly, even the spatial distribution of novae in M87 in general (Shafter et~al.\ 2000) appears to show such a tendency, albeit on larger scales (Fig.~3).

One source, observed in 1995, was found to be exceedingly luminous ($U=21.0$). This source is brighter by about two magnitudes than the brightest nova observed in M31 \citep{capaccioli89}. \citet{livio00} suggested that off-center helium ignitions may produce a population of ``super novae,'' characterized by $\sim0.15$~M$_{\odot}$ of Ni and He.  SN~1885A in M31 is the only event so far suspected to be such a ``super nova.'' 

We can exclude the event being a supernova.  The observation of the luminous transient was on 5~July 1995.  The previous observations were WFPC2 images obtained 27~May 1995 and the following observation was the WFPC2 image set obtained 23~November 1995, i.e.\ 38 days earlier and 141 days later.  Supernovae in elliptical galaxies are of Type~Ia which at peak have an absolute magnitude $M_U \sim -19.5$, \citep{wells94,schaefer95} or at Virgo distances, $U\sim 11.5$.  If the event with $U\approx 21$ was a supernova in decline, 38~days earlier it would have been much brighter, and trivially visible on the WFPC images (which are blank both before and after). Hence, it must be ``on the way up.''  Even so, 141~days after maximum of a Type~Ia supernova, the event would have declined from peak luminosity by only about 5~mag, so we would expect a source of around 16--17~mag in the November 1995 data. That is very much brighter than the globular clusters that are easily seen and so we can exclude the option of an undetected supernova. 

\section{Identifying the Progenitors of SNe Ia}

The measurement of potentially enhanced classical nova and SN~Ia rates in the vicinity of the jets of AGN, if confirmed by future studies, would provide the first means to date to determine if SNe~Ia are indeed formed by gradual accretion. Such an observation would also, finally, identify the progenitors of SNe~Ia. Theoretical models propose two possible candidates for the progenitor systems 
\citep{livio01,webbink84,whelan73,branch95}: (i)~close, double white dwarf systems in which the white dwarfs coalesce due to the emission of gravitational wave radiation, and (ii)~systems like supersoft X-ray sources or symbiotic stars, in which a single white dwarf accretes mass from a non-degenerate companion. Since the jet would simply act to enhance accretion (as traced by classical novae and SNe~Ia), we could conclude that the gradual accretion onto a white dwarf in \textit{single-degenerate systems} is a viable option ((ii) above).

The following natural question arises from the consideration of the preceding diagnostic of SN~Ia formation: is such an enhanced circumnuclear rate suggested by the current sample of a few hundred SNe~Ia? Unfortunately, such data are not available. SN hunters have long known that their tools limited their ability to detect SNe near galactic nuclei. The Shaw effect describes the loss of sensitivity of photographic plates near the centers of saturated, bright nuclei. In addition, the blurring effects of the atmosphere limit a distinction between nucleus variability and a SN to an arcsecond in angular separation. Space-based observatories like HST can discriminate SNe far closer to the nucleus. The SN~Ia 2000fs was discovered only $0.5''$ from the nucleus. Two of the other handful of SNe found by HST were found at $0.1''$ from the host nucleus \citep{gilliland97,beckwith01}. Future studies from space are required to expand this still meager sample. Upcoming STIS observations of M87 should address the equivalent question regarding the rates of novae.

On an even more speculative note, the suggested relation between AGN jets and enhanced rates of novae and SNe~Ia (if confirmed) can also have interesting and identifiable consequences for the metallicity of the host galaxy. Since novae (especially of the O-Ne type) have metallicities that are a few tens that of the sun \citep{gehrz98}, and SNe~Ia represent the major source of Fe-group nuclei \citep{thielemann01}, one prediction of a jet-induced rate enhancement is the generation of an \textit{azimuthal} metallicity gradient in the galaxy. More generally, jets may be expected to enhance the metallicity in AGN cores. 

\acknowledgements
We are grateful to Andr\'e Martel for providing us with helpful information, and to an anonymous referee for useful comments.

\clearpage
\begin{figure}
\plotone{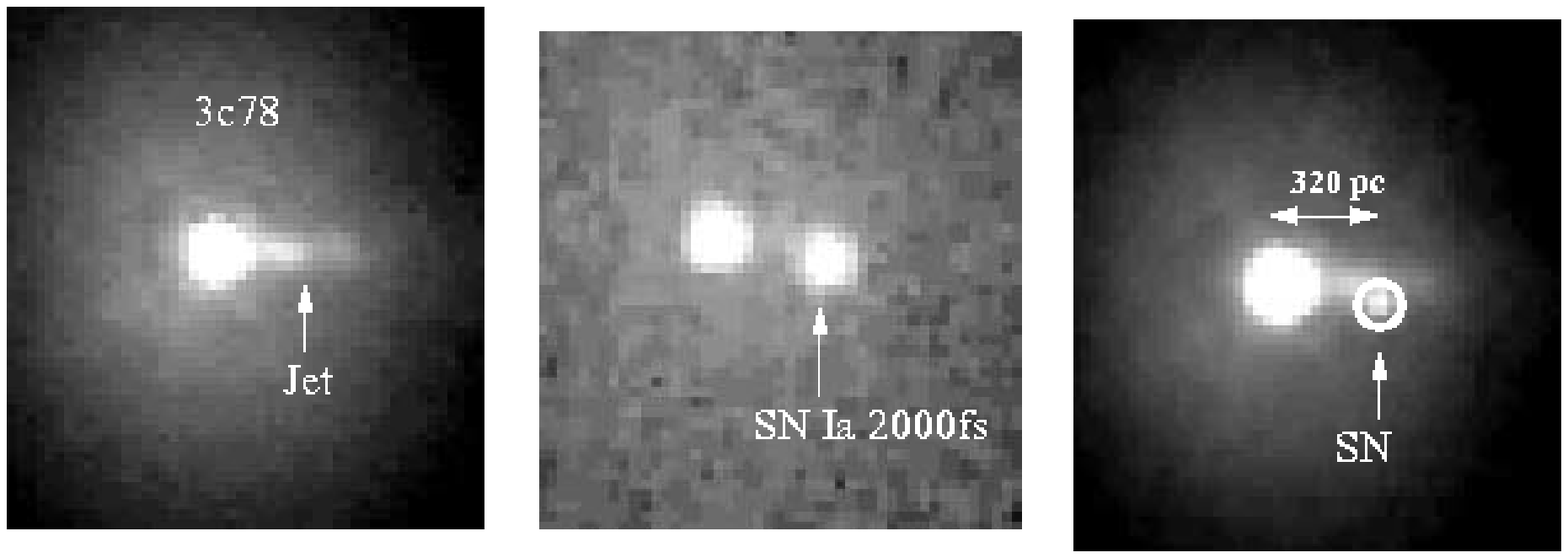}
\caption{Observations of SN~Ia 2000fs in close proximity to the optical jet in the host galaxy, 3C~78.  The panel on the left shows the galactic nucleus and the jet in a 1000 second image from HST before the SN was detected.  The middle panel shows the HST STIS acquisition image of the galactic nucleus serendipitiously containing the SN in the $5''$ by $5''$ field of view.  The image on the right shows a superposition of the preceding two images with the SN scaled down in flux to better shows its position relative to the jet.}
\end{figure}

\begin{figure}
\plotone{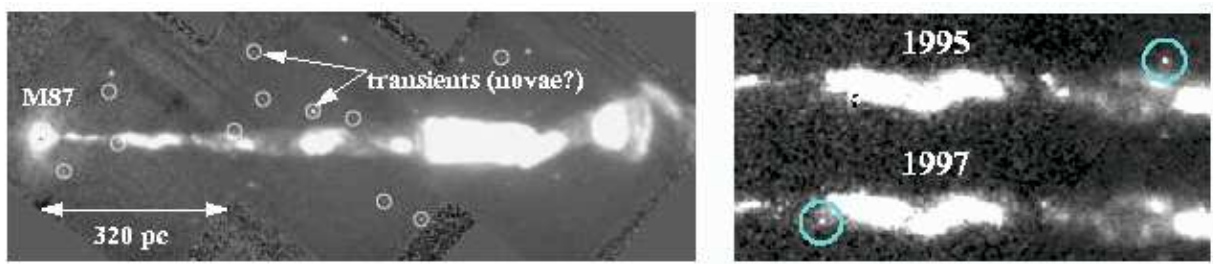}
\caption{Observations of the M87 jet using the Faint Object Camera on board the Hubble Space Telescope over a period of five years reveal the presence of a large number of transient sources, circled. Fig.~2 (a)~shows the locations of eleven secure identifications, and Fig.~2 (b)~shows, in more detail, the two closest to the jet.}
\end{figure}

\begin{figure}
\epsscale{.8}
\plotone{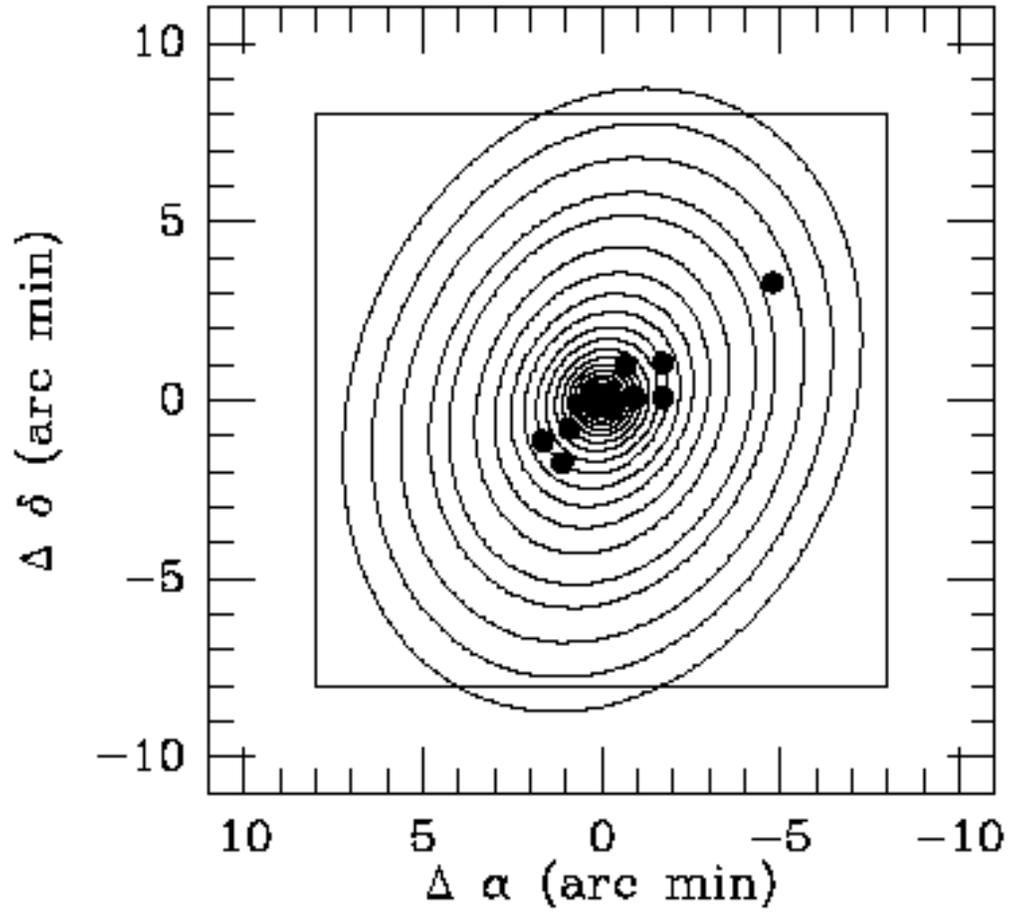}
\caption{An indication of a potential jet-novae relationship (jet p.a.~= 293~degrees). From Shafter et~al.\ (2000).}
\end{figure}
\end{document}